\documentclass[reviewcopy]{elsarticle}

\usepackage[reviewcopy]{adndt}
\usepackage{longtable}

\usepackage{amsmath}

\usepackage{amssymb}

\biboptions{square,sort&compress}
\bibpunct[]{[}{]}{,}{n}{}{;}
\begin{document}

\begin{frontmatter}

\journal{Atomic Data and Nuclear Data Tables}


\title{Radiative rates for E1, E2, M1, and M2 transitions in  F-like ions with 12 $\le$ Z $\le$ 23}

  \author[One]{Kanti M. Aggarwal\fnref{}\corref{cor1}}
  \ead{K.Aggarwal@qub.ac.uk}



  \cortext[cor1]{Corresponding author.}

  \address[One]{Astrophysics Research Centre, School of Mathematics and Physics, Queen's University Belfast,\\Belfast BT7 1NN,
Northern Ireland, UK}


\date{16.12.2002} 

\begin{abstract}  
In this paper, energy levels, radiative rates and lifetimes are reported for 11 F-like ions with 12 $\le$ Z $\le$ 23. Up to 198 levels (depending on the ion) have been considered which include 113  of the 2s$^2$2p$^5$, 2s2p$^6$, 2s$^2$2p$^4$3$\ell$, 2s2p$^5$3$\ell$, and 2p$^6$3$\ell$ configurations.   The general-purpose relativistic atomic structure package ({\sc grasp})  and the flexible atomic code ({\sc fac})    have been adopted for calculating the energy levels, but the {\sc grasp} alone for the remaining parameters.   Radiative rates (along with   oscillator strengths and line strengths) are listed  for all E1, E2, M1, and M2 transitions of the ions. Comparisons are made with  earlier available theoretical and experimental energies, for all ions, in order to assess the accuracy of the calculations. Comparisons have also been made for the radiative rates and lifetimes, which have been possible for only those among the lowest 60 levels.  \\ \\

{\em Received}: 22 November 2018, {\em Accepted}: 2 January  2019

\vspace{0.5 cm}
\end{abstract}

\end{frontmatter}




\newpage

\tableofcontents
\listofDtables
\listofDfigures
\vskip5pc


\section{Introduction}

F-like ions have been of interest for the diagnostics and modelling of astrophysical and fusion plasmas for which information about atomic data are required, particularly for energy levels, radiative rates  and collision strengths. The first major study for these ions was undertaken by Sampson et al. \cite{samp}, who performed calculations for a wide range of F-like ions with 22 $\le$ Z $\le$ 92 by using their Dirac-Fock-Slater (DFS) code. However, to economise space they reported only limited results, and for only a few ions, for energy levels, oscillator strengths (f-values) and collision strengths ($\Omega$). Since these data are too limited for applications, several workers after them have performed (more accurate) calculations for a variety of atomic parameters, but only for a section of these ions. For example, during the past one decade  in a series of papers \cite{kr,xe,w66a,w66b,w66c,ak1,ak2}, we have reported energy levels, radiative rates (A-values), oscillator strengths (f-values), line strengths (S-values), and lifetimes ($\tau$) for  F-like ions with 36 $\le$ Z $\le$ 74.  Similarly, Si et al. \cite{si} and Li et al. \cite{li} have reported data for ions with 24 $\le$ Z $\le$ 30 and 31 $\le$ Z $\le$ 35, respectively. Since their data are of comparable high accuracy there is no (real) need to revisit these ions. Therefore, in this paper we list our results for further 11 ions with 12 $\le$ Z $\le$ 23.  

As in the past, for our calculations we employ the general-purpose relativistic atomic structure package ({\sc grasp}) code \cite{grasp0}. However, this original version (referred to as GRASP0) has been extensively modified by (one of the authors) P.~H.~Norrington, and is available at the website  {\tt http://amdpp.phys.strath.ac.uk/UK\_APAP/codes.html}. Since the ions considered in this paper are comparatively lighter, a much larger CI (configuration interaction) has been considered than in the earlier works \cite{kr,xe,w66a,w66b,w66c,ak1,ak2}. In fact, for ions towards the neutral end a much (much) larger CI is required to improve accuracy of calculations and to obtain a better match with measurements, particularly for energy levels. To achieve this aim, calculations have also been performed with the flexible atomic code (FAC) of Gu \cite{fac}.

Earlier work for F-like ions with 12 $\le$ Z $\le$ 23 has mainly been performed by Gu \cite{gu} and J\"{o}nsson et al. \cite{jag}. Gu \cite{gu}  combined CI with many-body perturbation theory (MBPT) approach in FAC and  calculated energies, but only for the lowest three levels. For the same three levels, later on J\"{o}nsson et al. \cite{jag} reported energies and A-values for  which they adopted the revised version of the GRASP code \cite{grasp2k}, but  included a very large CI in the calculations. In spite of the high accuracy of their calculations, the reported data are too limited for applications. Nevertheless,  Froese Fischer and Tachiev \cite{cff} have already reported a larger amount of data among 60 levels of the 2s$^2$2p$^5$, 2s2p$^6$ and  2s$^2$2p$^4$3$\ell$ configurations of F-like ions with 9 $\le$ Z $\le$ 22, for which they adopted the multi-configuration Hartree-Fock (MCHF) code, included relativistic effects through Breit-Pauli Hamiltonian and considered a  large CI. Their calculations are (probably) the most accurate available todate because discrepancies with the measured energy levels are insignificant. Therefore, our aim is not to improve upon their work but to extend it to higher levels, particularly those of the 2s2p$^5$3$\ell$ and 2p$^6$3$\ell$ configurations. 

\section{Energy levels}

In our calculations with GRASP, for the optimisation of the orbitals we have used the option of  `extended average level' (EAL),  in which a weighted (proportional to 2$j$+1) trace of the Hamiltonian matrix is minimised.  The contributions of Breit and quantum electrodynamic effects (QED) are also included, although these are (comparatively) more  important for the heavier ions.  Calculations for energy levels and subsequent other parameters have been performed for up to 833 levels of 62 configurations, namely  2s$^2$2p$^5$, 2s2p$^6$, 2s$^2$2p$^4n\ell$, 2s2p$^5n\ell$, and  2p$^6n\ell$, where $n \le$ 7 and $\ell \le$ f for $n$ = 6 and 7.  In our earlier works \cite{kr,xe,w66a,w66b,w66c,ak1,ak2} only 38 configurations up to $n$ = 5 were considered which gave rise to a maximum of 501 levels, and were sufficient for accurate determination of energy levels. Therefore, the present work is a significant enhancement over the earlier one(s) for heavier ions, and further increase of CI is not possible with our version of the code.

As has already been stated that CI is very important for the ions under consideration here, we have therefore performed analogous calculations with FAC. This code is freely available at the website {\tt https://www-amdis.iaea.org/FAC/},  is  fully relativistic as others are, and in most instances yields energy levels of comparable accuracy. Additionally, it has an advantage of  efficiency and a (very) very  large CI can be included with ease in the calculations. Therefore, we have performed a series of calculations by gradually increasing CI, and our {\em final} one includes 1~70~649 levels (or configuration state functions, CSF) arising from all possible combinations of the 2*7, 2*6 n*1 (n $\le$ 13), 2*5 n*2 (n $\le$ 5), 2*5 3*1 n*1 (n $\le$ 13), and 2*5 4*1 5*1 configurations. These large calculations have fully stretched our computational resources and are significantly larger than those considered in earlier works, which included only up to 72~259 CSFs of the $n \le$ 5 configurations. 

Our earlier works  on F-like ions  \cite{kr,xe,w66a,w66b,w66c,ak1,ak2} focussed  on 113 levels of the 2s$^2$2p$^5$, 2s2p$^6$, 2s$^2$2p$^4$3$\ell$, 2s2p$^5$3$\ell$, and 2p$^6$3$\ell$ (11) configurations. However, for the heavier ions considered earlier, in most cases these levels were the {\em lowest} in energy whereas for the present ones there is a mixing with those from others, particularly of 2s$^2$2p$^4$4$\ell$. For this reason, the number of levels considered vary from ion to ion (121 for Mg~IV but 198 for V~XV), and the suitable cut-off implemented is the stage at which levels of the $n$ = 5 configurations start appearing.  

In Tables 1--11 we list our energy levels from both GRASP and FAC calculations along with the measured values assessed, compiled and recommended by the NIST (National Institute of Standards and Technology) team \cite{team}, and also available at  the  website {\tt http://www.nist.gov/pml/data/asd.cfm}.  Also included in these tables are the results from  the MCHF calculations of Froese Fischer and Tachiev \cite{cff}, but only for the lowest 60 levels. It may be noted that the results for Sc~XIII are {\em not} included here (or in subsequent tables), because these have already been discussed in a separate paper \cite{sc13}. Finally, as already stated and demonstrated in earlier works, with the inclusion of same CI there are no appreciable discrepancies in energies (for most of the levels) obtained from the GRASP and FAC codes. Therefore, whatever the differences in energies from these two codes we notice in these tables are mainly because of the distinctly different levels of CI included.

Before we discuss our results in detail, we would like to emphasize that the level designations provided in Tables~1--11 may not always be unambiguous, because about a dozen  levels are highly mixed for each ion. Often mixing coefficient from a particular level/configuration  dominates in two levels. Although this is a general atomic structure problem, irrespective of the code adopted, we have tried to provide a unique designation for each level, but that may change with other calculations and/or workers. Therefore,  only the $J^{\pi}$ values should be considered to be (more) definitive.

\subsection{Mg IV}

NIST energies are available for many levels, including the higher ones, but not all, whereas the MCHF calculations of Froese Fischer and Tachiev \cite{cff} are for the lowest 58 alone. Orderings in both GRASP and MCHF calculations is compatible for most levels, and some minor differences are for only a few, such as 39--45 in Table~1. However, for almost all levels the orderings are compatible between NIST and MCHF, and practically there are no discrepancies in magnitude either. On the other hand, our energies obtained from the GRASP code differ by up to 0.16~Ryd, for most levels. Only for level 3 (2s2p$^6$~$^2$S$_{1/2}$) the GRASP energy is higher, but is lower than all others listed in Table~1. Although such differences amount to a maximum of $\sim$5\%, there is a clear scope for improvement, particularly when the MCHF energies fully match with those of NIST. Since we have limitations with the GRASP code, our FAC energies (obtained with much larger CI) are clearly more accurate, and the discrepancies with those of NIST or MCHF are negligible for most levels, except for 3 (3, 29 and 30) for which the differences are below 0.1~Ryd. Based on the comparisons shown in Table~1 and discussed above, we may confidently state that our energies from FAC for all levels are as accurate as those of MCHF are for the lower ones, and can therefore be reliably employed in any modelling application.

\subsection{Al V to K XI}

For these ions the analysis of results and the conclusions drawn are similar to those for Mg~IV, i.e. the MCHF energies are closest to those of NIST and there are no major discrepancies with those from FAC, but the ones from GRASP are comparatively less accurate. However, there are some minor differences also. For example, the MCHF energies are not always for the lowest levels because there is a mixing from higher ones -- see for example, levels 59 to 61 in Table~2 for Al~V. Similarly, in a few instances the orderings of the MCHF energies slightly differ with those of NIST -- see for example, 20--22 in Table~2. Finally, the number of levels for which the NIST energies are available for these ions are often much less(er) than those for Mg~IV.

\subsection{Ca XII and Ti XIV}

For these two ions also the differences and/or similarities among different sets of energies listed in Tables 9 and 10 are comparable with those for others in section 2.2. However, for these two ions there is an unexpected anomaly in the listing of levels provided by Froese Fischer and Tachiev \cite{cff}. They have listed their energies for the 2s$^2$2p$^4$($^1$D$_2$)4s~$^2$D$_{5/2}$ level at 28.8057 and 37.5209~Ryd, for the respective two ions. For this level there is no correspondence with our calculations with both GRASP (38.3884 and 50.0932 Ryd) and FAC (38.5079 and 50.2149 Ryd). Furthermore, their calculations {\em do not} include levels of the 2s$^2$2p$^4$4s configuration for any F-like ion, and in fact their listed energies are closer to that of the 2s$^2$2p$^4$($^1$D$_2$)3s~$^2$D$_{5/2}$ level at 28.7969 and 37.4994~Ryd, respectively. Therefore, it is just a (typing) mistake in their table~6 for these two ions.

\subsection{V XV}
For this ion, Froese Fischer and Tachiev \cite{cff} have not calculated energies, but  Jup\'{e}n  et al.  \cite{jup} have with the Hartree-Fock relativistic (HFR) code of Cowan \cite{cow}, and therefore their results are included in Table~11. Their calculated energies are for 70 levels, but these are not the lowest and have quite a large spread. Furthermore, some of their results are based on interpolation, not actual calculations. Nevertheless, there are no appreciable discrepancies between our FAC and their HFR energies, although the GRASP ones are, as for other ions, comparatively less accurate and differ by up to $\sim$0.16~Ryd.  Apart from this, two levels (38 and 45, i.e. (2s$^2$2p$^4$($^3$P)3d) $^2$F$_{7/2}$ and $^4$F$_{7/2}$) are interchanged between our and their calculations. However, these two levels are {\em mixed} with the coefficients 0.721~$^2$F$_{7/2}$ + 0.614~$^4$F$_{7/2}$ and 0.689~$^4$F$_{7/2}$ + 0.609~$^2$F$_{7/2}$, respectively, in our calculations with GRASP, and in percentage terms are 52\%~$^2$F$_{7/2}$ + 38\%~$^4$F$_{7/2}$ and 47\%~$^4$F$_{7/2}$ + 37\%~$^2$F$_{7/2}$, respectively, whereas in the HFR are 94\%~$^2$F$_{7/2}$ and 54\%~$^4$F$_{7/2}$, respectively. Therefore, these two (and a few more) levels are inter-changeable in designation, as already stated. 
 
 \section{Radiative rates}\label{sec.eqs} 
 
 Apart from energy levels, data are required for A-values, and preferably  for  four types of transitions, namely electric dipole (E1),  magnetic dipole (M1), electric quadrupole (E2), and magnetic quadrupole (M2). These results are useful for the further calculations of lifetimes as well as for plasma modelling.  Our calculated results  with the {\sc grasp} code are listed  in Tables 12--22 for transition energies (wavelengths, $\lambda_{ji}$ in ${\rm \AA}$), radiative rates (A-values, in s$^{-1}$), oscillator strengths (f-values, dimensionless), and line strengths (S-values, in atomic units, 1 a.u. = 6.460$\times$10$^{-36}$ cm$^2$ esu$^2$)  for E1 transitions in F-like ions with 12 $\le$ Z $\le$ 23.  These results have been obtained in two gauges, namely  velocity  and length or Coulomb and Babushkin, respectively. In a perfectly accurate calculation  both forms (or gauges) are expected to provide  comparable results with their  ratio (R) being close(r) to unity. However, in reality this mostly applies to strong allowed transitions and the weak(er) ones with small f-values may differ substantially.  Therefore, we have also listed R in these tables for all E1 transitions, which are not only dominant but also  the most important in any calculation.  Corresponding results for  E2, M1 and M2 transitions are also listed, but  only for the A-values, because  data for f- or S-values can be easily   obtained using Eqs. (1-5) given in \cite{kr}.   The  indices used in these tables to represent the lower and upper levels of a transition correspond to those defined in Tables 1--11. Furthermore,  for brevity only transitions from the lowest 3 to higher excited levels are listed in Tables 12--22, but  full tables in the ASCII format are available online in the electronic version.

In Table~A we compare our A-values from GRASP with those of Froese Fischer and Tachiev \cite{cff}  with MCHF for some E1 transitions (from lowest 3 to up to  J = 30) for three ions, namely Mg~IV, Cl~IX and Ti~XIV. Also included in this table are our f-values from GRASP to give an idea about the strength of a transition. This comparison, although limited, will give sufficient idea about the (dis)agreements between the two calculations and hence some assessment of accuracy. Most of these transitions are weak with f $<$ 0.1 and therefore the differences between the two independent calculations are up to a factor of two -- see for example the 3--13 transition of all ions. Weak transitions are highly variable in magnitude with differing amount of CI (and/or methods/codes) and such differences are therefore quite normal for any ion. However, there are three transitions (1--3, 1--7 and 2--3) of Mg~IV  which have comparatively larger (but not much) f-value of $\sim$0.1 and differences between the two calculations are of $\sim$50\% for two of these, i.e. 1--3 and 2--3. Again, such differences are not surprising as even for much stronger ones it happens with differing amount of CI and/or methods as demonstrated  and discussed earlier for Mg-like ions \cite{mglike}. With increasing amount of CI (from 113 to 833 CSFs) we have performed 6 calculations with GRASP and the variation in A-values for both of these transitions is within 10\%. Similarly, some adjustment in A-values can be made by replacing transition energies with the measured ones, as A $\propto~ \Delta$E$^3$. However, it makes a difference of only $\sim$15\% for both of these transitions.  For two of these ions, Cl~IX and Ti~XIV,  J\"{o}nsson et al. \cite{jag} have reported A-values which are 2.867$\times$10$^{10}$ and 1.321$\times$10$^{10}$~s$^{-1}$ for the 1--3 and 2--3 transitions of Cl~IX and 5.163$\times$10$^{10}$ and 2.136$\times$10$^{10}$~s$^{-1}$ for Ti~XIV, and agree closely with those of MCHF \cite{cff} and within 10\% with our results. Therefore, we have confidence in our data as there are no major discrepancies with the earlier works.

A criterion often used (but never fully satisfactory or conclusively reliable) to assess the accuracy of A-values is to compare R, the ratio of velocity and length forms of a transition. For Mg~IV, there are 181 E1 transitions with f $\ge$ 0.1, and most of these have R within 20\% of unity. There are only 10 transitions which have a larger R, but only up to 1.5, and some include those listed in Table~A. Therefore, we may state with confidence that a majority of (strong) E1 transitions listed  in Tables~12-22 are accurate to about 20\%. Another conclusion we may draw is that  some small differences between the theoretical and experimental energy levels do not greatly affect the accuracy of A-values. However, this conclusion is fairly well known and is not new. Finally, we discuss below the lifetimes which will throw more analysis about the accuracy of our A-values, but before that we discuss the other types of transitions, i.e. E2, M1 and M2.

For a majority of transitions for any ion the f-values for E2, M1 and M2 transitions are generally (several orders of magnitude) smaller than for E1, but are desirable for considering a complete model in an analysis as well as for the calculations of lifetimes. Very limited data for such transitions are provided by Froese Fischer and Tachiev \cite{cff} and J\"{o}nsson et al. \cite{jag}, but in Table~B we make comparisons for  1--2 M1 and 1--2 E2 which are {\em common} among all calculations. As expected and also noted for other F-like ions earlier, there are no discrepancies among different sets of A-values, and this provides further confidence in the assessed accuracy of our listed results.

\section{Lifetimes}

The lifetime $\tau$ (s) of a level $j$ is calculated  as  1.0/$\Sigma_{i}$A$_{ji}$ where the summation runs over all types of transitions, although the E1 are normally the most dominant in magnitude, and hence  more important in its determination. Since $\tau$ is a measurable quantity, it provides a check on the accuracy of calculations, but no experiments have yet been performed for transitions/levels of F-like ions of present interest.  However, Froese Fischer and Tachiev \cite{cff}  have reported $\tau$ and therefore  in Tables 1--11 we have included their results along with ours to facilitate a direct comparison.  For most levels there are no discrepancies between the two calculations and the differences (if any) are within $\sim$20\%. However, for a few levels the discrepancies are rather large, up to 50\%, see for example levels 29/30 and 41/43 of Mg~IV in Table~1. These differences are a direct consequence of the corresponding differences in the A-values between the two calculations. For example, for level 29 (2s$^2$2p$^4$($^1$D)3p~$^2$P$^o_{3/2 }$) the major contributions are made by four transitions, i.e. from levels 3, 7, 8, and 9 with our A-values being 3.15, 8.72, 1.78, and 7.24 (10$^8$ s$^{-1}$), respectively, whereas those of MCHF \cite{cff} are 1.48, 5.71, 1.19, and 6.17 (10$^8$ s$^{-1}$), respectively. Needless to say, none of these transitions is strong as  their corresponding f-values are 0.0096, 0.1229, 0.0518, and 0.1384, respectively. Therefore, as for the A-values  our assessment of accuracy for $\tau$ is also the same, i.e. $\sim$20\%, for most levels.

\section{Conclusions}

In this paper, energies (for a maximum number of 198 levels) for  11 F-like ions with 12 $\le$ Z $\le$ 23 are reported, which include 113 levels of the 2s$^2$2p$^5$, 2s2p$^6$, 2s$^2$2p$^4$3$\ell$, 2s2p$^5$3$\ell$, and 2p$^6$3$\ell$ configurations. An extensive CI with up to 1~70~649 CSFs has been included in the FAC code to obtain energies as accurately as possible, although calculations have also been performed with the GRASP code, but with limited CI of up to 833 CSFs. In comparison, the GRASP energies are not as accurate as with FAC, because of the limitations of CI, although all differences are below $\sim$0.15~Ryd.  For many levels, particularly the lowest 60, prior theoretical as well as experimental energies are also available with which to compare, and there are no (significant) discrepancies in magnitude with our current results, which have been produced for a larger number of levels. However, for a few levels there are minor differences in their orderings and similarly, for a few of them in each ion there may be ambiguity in their LSJ$^{\pi}$ designations. 

Radiative rates for four types of transitions, i.e. E1, E2, M1, and M2,  are also reported for all ions, although the E1 are the most important because of their dominance in magnitude.   Comparisons with the existing literature are possible for accuracy assessment, but are limited to  transitions among the lowest 60 levels alone.  However, there are no (major) discrepancies between  our and earlier results, and all A-values for significantly strong transitions are assessed to be accurate to about 20\%. Adopting the calculated A-values,  lifetimes  have also been determined (to an estimated accuracy of 20\%), and these compare well with the earlier results of Froese Fischer and Tachiev \cite{cff}.

Combined with our earlier results \cite{kr,xe,w66a,w66b,w66c,ak1,ak2, sc13}, along with those of Si et al. \cite{si} and Li et al. \cite{li}, the present work completes data for all F-like ions with Z $\le$ 74.  These works include a larger number of levels than generally available in the literature, are comparatively more accurate, and hence can be confidently and reliably applied in the diagnostics and modelling of a variety of plasmas, including astrophysical and fusion.




\begin{appendix}

\def\thesection{} 

\section{Appendix A. Supplementary data}

Owing to space limitations, only parts of Tables 12-22  are presented here, but full tables are being made available as supplemental material in conjunction with the electronic publication of this work. Supplementary data associated with this article can be found, in the online version, at doi:nn.nnnn/j.adt.2019.nn.nnn.

\end{appendix}



\clearpage
\newpage
\renewcommand{\baselinestretch}{1.0}
\footnotesize
\begin{longtable}{@{\extracolsep\fill}lllllllllllllllllllllll@{}}
\caption{Comparison between our present GRASP (A1 and f) and earlier MCHF (A2: \cite{cff})   A-values (s$^{-1}$) for some  E1 transitions of Mg~IV, Cl~IX and Ti~XIV. $a{\pm}b \equiv a{\times}$10$^{{\pm}b}$.} 
Trans. & &  \multicolumn{3}{c}{Mg~IV}  & \multicolumn{2}{l}{Trans.} & \multicolumn{3}{c}{Cl~IX} & \multicolumn{2}{l}{Trans.} & \multicolumn{3}{c}{Ti~XIV}   \\ \\
\hline \\
I  & J & A1 & f & A2 & I & J & A1 & f & A2 & I & J & A1 & f & A2  \\  \\
\hline \\
\endfirsthead\\
\caption[]{(continued)}
Trans. & &  \multicolumn{3}{c}{Mg~IV}  & \multicolumn{2}{l}{Trans.} & \multicolumn{3}{c}{Cl~IX} & \multicolumn{2}{l}{Trans.} & \multicolumn{3}{c}{Ti~XIV}   \\ \\
\hline \\
I  & J & A1 & f & A2 & I & J & A1 & f & A2 & I & J & A1 & f & A2  \\  \\
\hline \\

\endhead 
      1 &   3	&  1.632$+$10 &  1.135$-$01 &  1.132$+$10  &   1 &   3 & 3.292$+$10 &  7.644$-$02 &  2.864$+$10   &  1   &   3   & 5.636$+$10 &  6.070$-$02  & 5.212$+$10     \\
      1 &   4	&  7.284$+$06 &  5.871$-$05 &  7.899$+$06  &   1 &   4 & 8.872$+$08 &  5.659$-$04 &  9.211$+$08   &  1   &   4   & 1.654$+$10 &  2.344$-$03  & 1.706$+$10     \\
      1 &   5	&  1.268$+$08 &  6.775$-$04 &  1.326$+$08  &   1 &   5 & 1.658$+$10 &  6.994$-$03 &  1.840$+$10   &  1   &   5   & 4.402$+$11 &  4.117$-$02  & 4.658$+$11     \\
      1 &   6	&  2.252$+$06 &  6.001$-$06 &  2.401$+$06  &   1 &   6 & 2.004$+$06 &  4.204$-$07 &  1.014$+$07   &  1   &   6   & 5.594$+$09 &  2.589$-$04  & 5.094$+$09     \\
      1 &   7	&  2.221$+$10 &  1.148$-$01 &  2.091$+$10  &   1 &   7 & 2.332$+$11 &  9.668$-$02 &  2.236$+$11   &  1   &   7   & 6.162$+$11 &  5.679$-$02  & 5.675$+$11     \\
      1 &   8	&  9.060$+$09 &  2.327$-$02 &  8.526$+$09  &   1 &   8 & 1.074$+$11 &  2.205$-$02 &  1.035$+$11   &  1   &   8   & 5.146$+$11 &  2.346$-$02  & 5.011$+$11     \\
      1 &   9	&  1.032$+$10 &  7.157$-$02 &  9.662$+$09  &   1 &   9 & 1.108$+$11 &  6.547$-$02 &  1.054$+$11   &  1   &   9   & 4.475$+$11 &  5.974$-$02  & 4.288$+$11     \\
      1 &  10	&  1.328$+$09 &  6.139$-$03 &  1.225$+$09  &   1 &  10 & 5.884$+$09 &  2.316$-$03 &  5.300$+$09   &  1   &  10   & 4.459$+$08 &  3.966$-$05  & 2.198$+$08     \\
      1 &  24	&  6.352$+$09 &  1.298$-$02 &  6.284$+$09  &   1 &  16 & 6.197$+$10 &  1.143$-$02 &  5.934$+$10   &  1   &  16   & 1.691$+$11 &  7.163$-$03  & 1.641$+$11     \\
      2 &   3	&  7.971$+$09 &  1.124$-$01 &  5.520$+$09  &   2 &   3 & 1.521$+$10 &  7.412$-$02 &  1.319$+$10   &  2   &   3   & 2.341$+$10 &  5.662$-$02  & 2.151$+$10     \\
      2 &   5	&  1.282$+$07 &  1.382$-$04 &  1.312$+$07  &   2 &   5 & 1.205$+$09 &  1.031$-$03 &  1.353$+$09   &  2   &   5   & 2.087$+$10 &  3.998$-$03  & 2.217$+$10     \\
      2 &   6	&  3.478$+$07 &  1.868$-$04 &  3.549$+$07  &   2 &   6 & 2.317$+$09 &  9.863$-$04 &  2.400$+$09   &  2   &   6   & 1.619$+$10 &  1.534$-$03  & 1.648$+$10     \\
      2 &   7	&  3.979$+$09 &  4.147$-$02 &  3.722$+$09  &   2 &   7 & 3.150$+$10 &  2.648$-$02 &  3.001$+$10   &  2   &   7   & 5.917$+$10 &  1.116$-$02  & 5.522$+$10     \\
      2 &   8	&  1.727$+$10 &  8.945$-$02 &  1.625$+$10  &   2 &   8 & 1.757$+$11 &  7.317$-$02 &  1.705$+$11   &  2   &   8   & 6.426$+$11 &  5.997$-$02  & 6.310$+$11     \\
      2 &  10	&  8.965$+$09 &  8.355$-$02 &  8.441$+$09  &   2 &  10 & 1.070$+$11 &  8.540$-$02 &  1.026$+$11   &  2   &  10   & 4.941$+$11 &  8.994$-$02  & 4.798$+$11     \\
      2 &  24	&  3.628$+$09 &  1.493$-$02 &  3.559$+$09  &   2 &  16 & 5.071$+$10 &  1.895$-$02 &  4.793$+$10   &  2   &  16   & 2.913$+$11 &  2.524$-$02  & 2.794$+$11     \\
      3 &  12	&  9.327$+$04 &  4.366$-$06 &  4.049$+$04  &   3 &  12 & 2.597$+$07 &  3.814$-$05 &  1.290$+$07   &  3   &  12   & 3.625$+$08 &  9.751$-$05  & 2.148$+$08     \\
      3 &  13	&  4.500$+$04 &  1.049$-$06 &  1.941$+$04  &   3 &  13 & 1.802$+$07 &  1.316$-$05 &  9.037$+$06   &  3   &  13   & 5.097$+$08 &  6.789$-$05  & 2.979$+$08     \\
      3 &  16	&  9.704$+$04 &  4.278$-$06 &  6.590$+$04  &   3 &  17 & 5.522$+$06 &  7.864$-$06 &  4.231$+$06   &  3   &  17   & 1.355$+$09 &  1.771$-$04  & 6.950$+$08     \\
      3 &  17	&  2.088$+$05 &  4.590$-$06 &  1.421$+$05  &   3 &  18 & 1.221$+$07 &  8.671$-$06 &  1.301$+$07   &  3   &  18   & 1.730$+$07 &  4.517$-$06  & 1.102$+$07     \\
      3 &  19	&  1.200$+$05 &  5.107$-$06 &  1.913$+$05  &   3 &  20 & 8.034$+$08 &  5.670$-$04 &  3.516$+$08   &  3   &  19   & 7.663$+$08 &  9.979$-$05  & 5.482$+$08     \\
      3 &  20	&  1.034$+$07 &  2.159$-$04 &  1.001$+$07  &   3 &  21 & 4.705$+$08 &  6.603$-$04 &  2.281$+$08   &  3   &  21   & 9.104$+$08 &  2.360$-$04  & 3.513$+$08     \\
      3 &  21	&  4.175$+$05 &  1.738$-$05 &  6.540$+$05  &   3 &  22 & 5.736$+$08 &  7.996$-$04 &  2.262$+$08   &  3   &  22   & 1.964$+$09 &  5.055$-$04  & 1.453$+$09     \\
      3 &  22	&  5.503$+$07 &  1.125$-$03 &  1.222$+$07  &   3 &  23 & 6.032$+$07 &  8.390$-$05 &  2.701$+$06   &  3   &  23   & 2.310$+$08 &  2.961$-$05  & 8.269$+$07     \\
      3 &  23	&  6.726$+$07 &  2.749$-$03 &  2.223$+$07  &   3 &  24 & 1.428$+$08 &  9.904$-$05 &  5.219$+$07   &  3   &  24   & 1.161$+$09 &  2.974$-$04  & 6.699$+$08     \\
      3 &  27	&  6.641$+$05 &  2.170$-$05 &  6.884$+$05  &   3 &  27 & 5.504$+$07 &  7.121$-$05 &  4.127$+$07   &  3   &  27   & 8.723$+$08 &  2.150$-$04  & 7.116$+$08     \\
      3 &  29	&  3.154$+$08 &  9.629$-$03 &  1.476$+$08  &   3 &  29 & 5.270$+$09 &  6.522$-$03 &  3.183$+$09   &  3   &  29   & 1.853$+$10 &  4.426$-$03  & 1.350$+$10     \\
      3 &  30	&  3.393$+$08 &  5.147$-$03 &  1.563$+$08  &   3 &  30 & 6.777$+$09 &  4.178$-$03 &  4.212$+$09   &  3   &  30   & 2.935$+$10 &  3.496$-$03  & 2.344$+$10     \\
\\  \hline            								                	 
\end{longtable}   								   					       
			      							   					       

														       
\begin{flushleft}													       
{\small
   
}															       
\end{flushleft} 



\renewcommand{\baselinestretch}{1.0}
\footnotesize
\begin{longtable}{@{\extracolsep\fill}lllllllll@{}}
\caption{Comparison of A-values (s$^{-1}$)  for the 1--2 M1 and 1--2 E2 transitions  of F-like ions with 12 $\le$ Z $\le$ 23. $a{\pm}b \equiv a{\times}$10$^{{\pm}b}$.}
Z  & \multicolumn{2}{c}{GRASP}    & \multicolumn{2}{c}{GRASP2K \cite{jag}}      &    \multicolumn{2}{c}{MCHF \cite{cff}} \\  \\
\hline \\
Transition & 1--2 M1 & 1--2 E2 & 1--2 M1 & 1--2 E2 & 1--2 M1 & 1--2 E2   \\
\\ \hline  \\  
\endfirsthead\\
\caption[]{(continued)}
Z  & \multicolumn{2}{c}{GRASP}    & \multicolumn{2}{c}{GRASP2K \cite{jag}}      &    \multicolumn{2}{c}{MCHF \cite{cff}} \\  \\
\hline \\
Transition & 1--2 M1 & 1--2 E2 & 1--2 M1 & 1--2 E2 & 1--2 M1 & 1--2 E2   \\
\\ \hline  \\  \\
\hline\\
\endhead 
12 & 1.788$-$01 & 6.461$-$07 &  	  &	       & 1.968$-$01 & 6.430$-$07  \\
13 & 6.780$-$01 & 3.400$-$06 &	 	  &	       & 7.299$-$01 & 3.397$-$06  \\
14 & 2.234$+$00 & 1.536$-$05 & 2.374$+$00 & 1.536$-$05 & 2.376$+$00 & 1.541$-$05  \\
15 & 6.582$+$00 & 6.098$-$05 & 6.910$+$00 & 6.094$-$05 & 6.954$+$00 & 6.155$-$05  \\
16 & 1.770$+$01 & 2.173$-$04 & 1.842$+$01 & 2.169$-$04 & 1.864$+$01 & 2.207$-$04  \\
17 & 4.408$+$01 & 7.055$-$04 & 4.559$+$01 & 7.037$-$04 & 4.637$-$01 & 7.219$-$04  \\
18 & 1.029$+$02 & 2.114$-$03 & 1.059$+$02 & 2.108$-$03 & 1.083$+$02 & 2.180$-$03  \\
19 & 2.269$+$02 & 5.909$-$03 & 2.326$+$02 & 5.888$-$03 & 2.392$+$02 & 6.144$-$03  \\
20 & 4.764$+$02 & 1.553$-$02 & 4.871$+$02 & 1.547$-$02 & 5.038$+$02 & 1.629$-$02  \\
22 & 1.855$+$03 & 9.162$-$02 & 1.888$+$03 & 9.124$-$02 & 1.976$-$03 & 9.793$-$02  \\
23 & 3.470$+$03 & 2.079$-$01 & 3.525$+$03 & 2.071$-$01 & 3.487$+$03$^a$ & 2.051$-$01$^a$  \\
\\  \hline            								                	 
\end{longtable}   								   					       
			      							   					       

														       
\begin{flushleft}													       
{\small
GRASP: present calculations with  the {\sc grasp} code for 833 levels \\
GRASP2K: earlier calculations of J\"{o}nsson  et al. \cite{jag}  with  the {\sc grasp2k} code  \\ 
MCHF: earlier calculations of  Froese Fischer and Tachiev \cite{cff} with  the {\sc mchf} code  \\   
$a$: These results are of Nandy and Sahoo \cite{ns} \\
}															       
\end{flushleft} 

\clearpage
\newpage


\TableExplanation

\bigskip
\renewcommand{\arraystretch}{1.0}

\section*{Table 1.\label{tbl1te} Energies (Ryd) for 121 levels of Mg~IV and their lifetimes ($\tau$, s).}
\begin{tabular}{@{}p{1in}p{6in}@{}}
Index            & Level Index \\
Configuration    & The configuration to which the level belongs \\
Level             & The $LSJ$ designation of the level \\
NIST              & Energies compiled by NIST and available at the website  {\tt http://www.nist.gov/pml/data/asd.cfm} \\
GRASP          & Present energies from the {\sc grasp} code  with 62  configurations and 833 level calculations \\
FAC             & Present energies from the {\sc fac} code  with 1~70~649 level calculations \\
MCHF          & Earlier calculations of Froese Fischer and Tachiev \cite{cff} with the {\sc mchf} code \\ 
$\tau$ (MCHF)       & Lifetime (in s) of the level  from the MCHF calculations of  Froese Fischer and Tachiev \cite{cff}  \\
$\tau$ (GRASP)       & Lifetime (in s) of the level  from present calculations with the {\sc grasp} code \\

\end{tabular}
\label{ExplTable1}

\bigskip
\renewcommand{\arraystretch}{1.0}

\section*{Table 2.\label{tbl2te} Energies (Ryd) for  125 levels of Al~V and their lifetimes ($\tau$, s). }
\begin{tabular}{@{}p{1in}p{6in}@{}}
Index            & Level Index \\
Configuration    & The configuration to which the level belongs \\
Level             & The $LSJ$ designation of the level \\
NIST              & Energies compiled by NIST and available at the website  {\tt http://www.nist.gov/pml/data/asd.cfm} \\
GRASP          & Present energies from the {\sc grasp} code  with 62  configurations and 833 level calculations \\
FAC             & Present energies from the {\sc fac} code  with 1~70~649 level calculations \\
MCHF          & Earlier calculations of Froese Fischer and Tachiev \cite{cff} with the {\sc mchf} code \\ 
$\tau$ (MCHF)       & Lifetime (in s) of the level  from the MCHF calculations of  Froese Fischer and Tachiev \cite{cff}  \\
$\tau$ (GRASP)       & Lifetime (in s) of the level  from present calculations with the {\sc grasp} code \\

\end{tabular}
\label{ExplTable2}

\bigskip
\renewcommand{\arraystretch}{1.0}

\section*{Table 3.\label{tbl3te} Energies (Ryd) for 137 levels of Si~VI and their lifetimes ($\tau$, s). }
\begin{tabular}{@{}p{1in}p{6in}@{}}
Index            & Level Index \\
Configuration    & The configuration to which the level belongs \\
Level             & The $LSJ$ designation of the level \\
NIST              & Energies compiled by NIST and available at the website  {\tt http://www.nist.gov/pml/data/asd.cfm} \\
GRASP          & Present energies from the {\sc grasp} code  with 62  configurations and 833 level calculations \\
FAC             & Present energies from the {\sc fac} code  with 1~70~649 level calculations \\
MCHF          & Earlier calculations of Froese Fischer and Tachiev \cite{cff} with the {\sc mchf} code \\ 
$\tau$ (MCHF)       & Lifetime (in s) of the level  from the MCHF calculations of  Froese Fischer and Tachiev \cite{cff}  \\
$\tau$ (GRASP)       & Lifetime (in s) of the level  from present calculations with the {\sc grasp} code \\

\end{tabular}
\label{ExplTable3}

\bigskip
\renewcommand{\arraystretch}{1.0}

\section*{Table 4.\label{tbl4te} Energies (Ryd) for 163 levels of P~VII and their lifetimes ($\tau$, s). }
\begin{tabular}{@{}p{1in}p{6in}@{}}
Index            & Level Index \\
Configuration    & The configuration to which the level belongs \\
Level             & The $LSJ$ designation of the level \\
NIST              & Energies compiled by NIST and available at the website  {\tt http://www.nist.gov/pml/data/asd.cfm} \\
GRASP          & Present energies from the {\sc grasp} code  with 62  configurations and 833 level calculations \\
FAC             & Present energies from the {\sc fac} code  with 1~70~649 level calculations \\
MCHF          & Earlier calculations of Froese Fischer and Tachiev \cite{cff} with the {\sc mchf} code \\ 
$\tau$ (MCHF)       & Lifetime (in s) of the level  from the MCHF calculations of  Froese Fischer and Tachiev \cite{cff}  \\
$\tau$ (GRASP)       & Lifetime (in s) of the level  from present calculations with the {\sc grasp} code \\

\end{tabular}
\label{ExplTable4}

\bigskip
\renewcommand{\arraystretch}{1.0}

\section*{Table 5.\label{tbl5te} Energies (Ryd) for  178  levels of S~VIII and their lifetimes ($\tau$, s). }
\begin{tabular}{@{}p{1in}p{6in}@{}}
Index            & Level Index \\
Configuration    & The configuration to which the level belongs \\
Level             & The $LSJ$ designation of the level \\
NIST              & Energies compiled by NIST and available at the website  {\tt http://www.nist.gov/pml/data/asd.cfm} \\
GRASP          & Present energies from the {\sc grasp} code  with 62  configurations and 833 level calculations \\
FAC             & Present energies from the {\sc fac} code  with 1~70~649 level calculations \\
MCHF          & Earlier calculations of Froese Fischer and Tachiev \cite{cff} with the {\sc mchf} code \\ 
$\tau$ (MCHF)       & Lifetime (in s) of the level  from the MCHF calculations of  Froese Fischer and Tachiev \cite{cff}  \\
$\tau$ (GRASP)       & Lifetime (in s) of the level  from present calculations with the {\sc grasp} code \\

\end{tabular}
\label{ExplTable5}

\bigskip
\renewcommand{\arraystretch}{1.0}

\section*{Table 6.\label{tbl6te} Energies (Ryd) for 187 levels of Cl~IX and their lifetimes ($\tau$, s). }
\begin{tabular}{@{}p{1in}p{6in}@{}}
Index            & Level Index \\
Configuration    & The configuration to which the level belongs \\
Level             & The $LSJ$ designation of the level \\
NIST              & Energies compiled by NIST and available at the website  {\tt http://www.nist.gov/pml/data/asd.cfm} \\
GRASP          & Present energies from the {\sc grasp} code  with 62  configurations and 833 level calculations \\
FAC             & Present energies from the {\sc fac} code  with 1~70~649 level calculations \\
MCHF          & Earlier calculations of Froese Fischer and Tachiev \cite{cff} with the {\sc mchf} code \\ 
$\tau$ (MCHF)       & Lifetime (in s) of the level  from the MCHF calculations of  Froese Fischer and Tachiev \cite{cff}  \\
$\tau$ (GRASP)       & Lifetime (in s) of the level  from present calculations with the {\sc grasp} code \\

\end{tabular}
\label{ExplTable6}

\bigskip
\renewcommand{\arraystretch}{1.0}

\section*{Table 7.\label{tbl7te} Energies (Ryd) for  193 levels of Ar~X and their lifetimes ($\tau$, s). }
\begin{tabular}{@{}p{1in}p{6in}@{}}
Index            & Level Index \\
Configuration    & The configuration to which the level belongs \\
Level             & The $LSJ$ designation of the level \\
NIST              & Energies compiled by NIST and available at the website  {\tt http://www.nist.gov/pml/data/asd.cfm} \\
GRASP          & Present energies from the {\sc grasp} code  with 62  configurations and 833 level calculations \\
FAC             & Present energies from the {\sc fac} code  with 1~70~649 level calculations \\
MCHF          & Earlier calculations of Froese Fischer and Tachiev \cite{cff} with the {\sc mchf} code \\ 
$\tau$ (MCHF)       & Lifetime (in s) of the level  from the MCHF calculations of  Froese Fischer and Tachiev \cite{cff}  \\
$\tau$ (GRASP)       & Lifetime (in s) of the level  from present calculations with the {\sc grasp} code \\

\end{tabular}
\label{ExplTable7}

\bigskip
\renewcommand{\arraystretch}{1.0}

\section*{Table 8.\label{tbl8te} Energies (Ryd) for  195 levels of K~XI and their lifetimes ($\tau$, s). }
\begin{tabular}{@{}p{1in}p{6in}@{}}
Index            & Level Index \\
Configuration    & The configuration to which the level belongs \\
Level             & The $LSJ$ designation of the level \\
NIST              & Energies compiled by NIST and available at the website  {\tt http://www.nist.gov/pml/data/asd.cfm} \\
GRASP          & Present energies from the {\sc grasp} code  with 62  configurations and 833 level calculations \\
FAC             & Present energies from the {\sc fac} code  with 1~70~649 level calculations \\
MCHF          & Earlier calculations of Froese Fischer and Tachiev \cite{cff} with the {\sc mchf} code \\ 
$\tau$ (MCHF)       & Lifetime (in s) of the level  from the MCHF calculations of  Froese Fischer and Tachiev \cite{cff}  \\
$\tau$ (GRASP)       & Lifetime (in s) of the level  from present calculations with the {\sc grasp} code \\

\end{tabular}
\label{ExplTable8}

\bigskip
\renewcommand{\arraystretch}{1.0}

\section*{Table 9.\label{tbl9te} Energies (Ryd) for 195 levels of Ca~XII and their lifetimes ($\tau$, s). }
\begin{tabular}{@{}p{1in}p{6in}@{}}
Index            & Level Index \\
Configuration    & The configuration to which the level belongs \\
Level             & The $LSJ$ designation of the level \\
NIST              & Energies compiled by NIST and available at the website  {\tt http://www.nist.gov/pml/data/asd.cfm} \\
GRASP          & Present energies from the {\sc grasp} code  with 62  configurations and 833 level calculations \\
FAC             & Present energies from the {\sc fac} code  with 1~70~649 level calculations \\
MCHF          & Earlier calculations of Froese Fischer and Tachiev \cite{cff} with the {\sc mchf} code \\ 
$\tau$ (MCHF)       & Lifetime (in s) of the level  from the MCHF calculations of  Froese Fischer and Tachiev \cite{cff}  \\
$\tau$ (GRASP)       & Lifetime (in s) of the level  from present calculations with the {\sc grasp} code \\

\end{tabular}
\label{ExplTable9}

\bigskip
\renewcommand{\arraystretch}{1.0}

\section*{Table 10.\label{tbl10te} Energies (Ryd) for  198 levels of Ti~XIV and their lifetimes ($\tau$, s).  }
\begin{tabular}{@{}p{1in}p{6in}@{}}
Index            & Level Index \\
Configuration    & The configuration to which the level belongs \\
Level             & The $LSJ$ designation of the level \\
NIST              & Energies compiled by NIST and available at the website  {\tt http://www.nist.gov/pml/data/asd.cfm} \\
GRASP          & Present energies from the {\sc grasp} code  with 62  configurations and 833 level calculations \\
FAC             & Present energies from the {\sc fac} code  with 1~70~649 level calculations \\
MCHF          & Earlier calculations of Froese Fischer and Tachiev \cite{cff} with the {\sc mchf} code \\ 
$\tau$ (MCHF)       & Lifetime (in s) of the level  from the MCHF calculations of  Froese Fischer and Tachiev \cite{cff}  \\
$\tau$ (GRASP)       & Lifetime (in s) of the level  from present calculations with the {\sc grasp} code \\

\end{tabular}
\label{ExplTable10}

\bigskip
\renewcommand{\arraystretch}{1.0}

\section*{Table 11.\label{tbl11te} Energies (Ryd) for 198 levels of V~XV and their lifetimes ($\tau$, s). }
\begin{tabular}{@{}p{1in}p{6in}@{}}
Index            & Level Index \\
Configuration    & The configuration to which the level belongs \\
Level             & The $LSJ$ designation of the level \\
NIST              & Energies compiled by NIST and available at the website  {\tt http://www.nist.gov/pml/data/asd.cfm} \\
GRASP          & Present energies from the {\sc grasp} code  with 62  configurations and 833 level calculations \\
FAC             & Present energies from the {\sc fac} code  with 1~70~649 level calculations \\
HFR          & Earlier calculations of Jup\'{e}n et al. \cite{jup} with the {\sc hfr} code \\ 
$\tau$ (GRASP)       & Lifetime (in s) of the level  from present calculations with the {\sc grasp} code \\

\end{tabular}
\label{ExplTable11}


\bigskip
\section*{Table 12.\label{tbl12te}  Transition wavelengths ($\lambda_{ij}$ in $\rm \AA$), radiative rates (A$_{ji}$ in s$^{-1}$),
 oscillator strengths (f$_{ij}$, dimensionless), and line strengths (S, in atomic units) for electric dipole (E1), and 
A$_{ji}$ for electric quadrupole (E2), magnetic dipole (M1), and magnetic quadrupole (M2) transitions of Mg~IV.
The ratio R(E1) of velocity and length forms of A-values for E1 transitions is listed in the last column.}
\begin{tabular}{@{}p{1in}p{6in}@{}}
$i$ and $j$         & The lower ($i$) and upper ($j$) levels of a transition as defined in Table 1.\\
$\lambda_{ij}$      & Transition wavelength (in ${\rm \AA}$) \\
A$^{E1}_{ji}$       & Radiative transition probability (in s$^{-1}$) for the E1 transitions \\
f$^{E1}_{ij}$       & Absorption oscillator strength (dimensionless) for the E1 transitions \\
S$^{E1}$            & Line strength in atomic unit (a.u.), 1 a.u. = 6.460$\times$10$^{-36}$ cm$^2$ esu$^2$ for the E1 transitions \\
A$^{E2}_{ji}$       & Radiative transition probability (in s$^{-1}$) for the E2 transitions \\
A$^{M1}_{ji}$       & Radiative transition probability (in s$^{-1}$) for the M1 transitions \\
A$^{M2}_{ji}$       & Radiative transition probability (in s$^{-1}$) for the M2 transitions \\
R(E1)                     & Ratio of velocity and length forms of A- (or f- and S-) values for the E1 transitions \\
$a{\pm}b$ &  $\equiv a\times{10^{{\pm}b}}$ \\
\end{tabular}
\label{ExplTable12}

\bigskip
\section*{Table 13.\label{tbl13te}  Transition wavelengths ($\lambda_{ij}$ in $\rm \AA$), radiative rates (A$_{ji}$ in s$^{-1}$),
 oscillator strengths (f$_{ij}$, dimensionless), and line strengths (S, in atomic units) for electric dipole (E1), and 
A$_{ji}$ for electric quadrupole (E2), magnetic dipole (M1), and magnetic quadrupole (M2) transitions of Al~V.
 The ratio R(E1) of velocity and length forms of A-values for E1 transitions is listed in the last column.}
\begin{tabular}{@{}p{1in}p{6in}@{}}
$i$ and $j$         & The lower ($i$) and upper ($j$) levels of a transition as defined in Table 2.\\
$\lambda_{ij}$      & Transition wavelength (in ${\rm \AA}$) \\
A$^{E1}_{ji}$       & Radiative transition probability (in s$^{-1}$) for the E1 transitions \\
f$^{E1}_{ij}$       & Absorption oscillator strength (dimensionless) for the E1 transitions \\
S$^{E1}$            & Line strength in atomic unit (a.u.), 1 a.u. = 6.460$\times$10$^{-36}$ cm$^2$ esu$^2$ for the E1 transitions \\
A$^{E2}_{ji}$       & Radiative transition probability (in s$^{-1}$) for the E2 transitions \\
A$^{M1}_{ji}$       & Radiative transition probability (in s$^{-1}$) for the M1 transitions \\
A$^{M2}_{ji}$       & Radiative transition probability (in s$^{-1}$) for the M2 transitions \\
R(E1)                     & Ratio of velocity and length forms of A- (or f- and S-) values for the E1 transitions \\
$a{\pm}b$ &  $\equiv a\times{10^{{\pm}b}}$ \\
\end{tabular}
\label{ExplTable13}

\bigskip
\section*{Table 14.\label{tbl14te}  Transition wavelengths ($\lambda_{ij}$ in $\rm \AA$), radiative rates (A$_{ji}$ in s$^{-1}$),
 oscillator strengths (f$_{ij}$, dimensionless), and line strengths (S, in atomic units) for electric dipole (E1), and 
A$_{ji}$ for electric quadrupole (E2), magnetic dipole (M1), and magnetic quadrupole (M2) transitions of Si~VI.
 The ratio R(E1) of velocity and length forms of A-values for E1 transitions is listed in the last column.}
\begin{tabular}{@{}p{1in}p{6in}@{}}
$i$ and $j$         & The lower ($i$) and upper ($j$) levels of a transition as defined in Table 3.\\
$\lambda_{ij}$      & Transition wavelength (in ${\rm \AA}$) \\
A$^{E1}_{ji}$       & Radiative transition probability (in s$^{-1}$) for the E1 transitions \\
f$^{E1}_{ij}$       & Absorption oscillator strength (dimensionless) for the E1 transitions \\
S$^{E1}$            & Line strength in atomic unit (a.u.), 1 a.u. = 6.460$\times$10$^{-36}$ cm$^2$ esu$^2$ for the E1 transitions \\
A$^{E2}_{ji}$       & Radiative transition probability (in s$^{-1}$) for the E2 transitions \\
A$^{M1}_{ji}$       & Radiative transition probability (in s$^{-1}$) for the M1 transitions \\
A$^{M2}_{ji}$       & Radiative transition probability (in s$^{-1}$) for the M2 transitions \\
R(E1)                     & Ratio of velocity and length forms of A- (or f- and S-) values for the E1 transitions \\
$a{\pm}b$ &  $\equiv a\times{10^{{\pm}b}}$ \\
\end{tabular}
\label{ExplTable14}

\bigskip
\section*{Table 15.\label{tbl15te}  Transition wavelengths ($\lambda_{ij}$ in $\rm \AA$), radiative rates (A$_{ji}$ in s$^{-1}$),
 oscillator strengths (f$_{ij}$, dimensionless), and line strengths (S, in atomic units) for electric dipole (E1), and 
A$_{ji}$ for electric quadrupole (E2), magnetic dipole (M1), and magnetic quadrupole (M2) transitions of P~VII
 The ratio R(E1) of velocity and length forms of A-values for E1 transitions is listed in the last column.}
\begin{tabular}{@{}p{1in}p{6in}@{}}
$i$ and $j$         & The lower ($i$) and upper ($j$) levels of a transition as defined in Table 4.\\
$\lambda_{ij}$      & Transition wavelength (in ${\rm \AA}$) \\
A$^{E1}_{ji}$       & Radiative transition probability (in s$^{-1}$) for the E1 transitions \\
f$^{E1}_{ij}$       & Absorption oscillator strength (dimensionless) for the E1 transitions \\
S$^{E1}$            & Line strength in atomic unit (a.u.), 1 a.u. = 6.460$\times$10$^{-36}$ cm$^2$ esu$^2$ for the E1 transitions \\
A$^{E2}_{ji}$       & Radiative transition probability (in s$^{-1}$) for the E2 transitions \\
A$^{M1}_{ji}$       & Radiative transition probability (in s$^{-1}$) for the M1 transitions \\
A$^{M2}_{ji}$       & Radiative transition probability (in s$^{-1}$) for the M2 transitions \\
R(E1)                     & Ratio of velocity and length forms of A- (or f- and S-) values for the E1 transitions \\
$a{\pm}b$ &  $\equiv a\times{10^{{\pm}b}}$ \\
\end{tabular}
\label{ExplTable15}

\bigskip
\section*{Table 16.\label{tbl16te}  Transition wavelengths ($\lambda_{ij}$ in $\rm \AA$), radiative rates (A$_{ji}$ in s$^{-1}$),
 oscillator strengths (f$_{ij}$, dimensionless), and line strengths (S, in atomic units) for electric dipole (E1), and 
A$_{ji}$ for electric quadrupole (E2), magnetic dipole (M1), and magnetic quadrupole (M2) transitions of S~VIII.
  The ratio R(E1) of velocity and length forms of A-values for E1 transitions is listed in the last column.}
\begin{tabular}{@{}p{1in}p{6in}@{}}
$i$ and $j$         & The lower ($i$) and upper ($j$) levels of a transition as defined in Table 5.\\
$\lambda_{ij}$      & Transition wavelength (in ${\rm \AA}$) \\
A$^{E1}_{ji}$       & Radiative transition probability (in s$^{-1}$) for the E1 transitions \\
f$^{E1}_{ij}$       & Absorption oscillator strength (dimensionless) for the E1 transitions \\
S$^{E1}$            & Line strength in atomic unit (a.u.), 1 a.u. = 6.460$\times$10$^{-36}$ cm$^2$ esu$^2$ for the E1 transitions \\
A$^{E2}_{ji}$       & Radiative transition probability (in s$^{-1}$) for the E2 transitions \\
A$^{M1}_{ji}$       & Radiative transition probability (in s$^{-1}$) for the M1 transitions \\
A$^{M2}_{ji}$       & Radiative transition probability (in s$^{-1}$) for the M2 transitions \\
R(E1)                     & Ratio of velocity and length forms of A- (or f- and S-) values for the E1 transitions \\
$a{\pm}b$ &  $\equiv a\times{10^{{\pm}b}}$ \\
\end{tabular}
\label{ExplTable16}

\bigskip
\section*{Table 17.\label{tbl17te}  Transition wavelengths ($\lambda_{ij}$ in $\rm \AA$), radiative rates (A$_{ji}$ in s$^{-1}$),
 oscillator strengths (f$_{ij}$, dimensionless), and line strengths (S, in atomic units) for electric dipole (E1), and 
A$_{ji}$ for electric quadrupole (E2), magnetic dipole (M1), and magnetic quadrupole (M2) transitions of Cl~IX.
  The ratio R(E1) of velocity and length forms of A-values for E1 transitions is listed in the last column.}
\begin{tabular}{@{}p{1in}p{6in}@{}}
$i$ and $j$         & The lower ($i$) and upper ($j$) levels of a transition as defined in Table 6.\\
$\lambda_{ij}$      & Transition wavelength (in ${\rm \AA}$) \\
A$^{E1}_{ji}$       & Radiative transition probability (in s$^{-1}$) for the E1 transitions \\
f$^{E1}_{ij}$       & Absorption oscillator strength (dimensionless) for the E1 transitions \\
S$^{E1}$            & Line strength in atomic unit (a.u.), 1 a.u. = 6.460$\times$10$^{-36}$ cm$^2$ esu$^2$ for the E1 transitions \\
A$^{E2}_{ji}$       & Radiative transition probability (in s$^{-1}$) for the E2 transitions \\
A$^{M1}_{ji}$       & Radiative transition probability (in s$^{-1}$) for the M1 transitions \\
A$^{M2}_{ji}$       & Radiative transition probability (in s$^{-1}$) for the M2 transitions \\
R(E1)                     & Ratio of velocity and length forms of A- (or f- and S-) values for the E1 transitions \\
$a{\pm}b$ &  $\equiv a\times{10^{{\pm}b}}$ \\
\end{tabular}
\label{ExplTable17}

\bigskip
\section*{Table 18.\label{tbl18te}  Transition wavelengths ($\lambda_{ij}$ in $\rm \AA$), radiative rates (A$_{ji}$ in s$^{-1}$),
 oscillator strengths (f$_{ij}$, dimensionless), and line strengths (S, in atomic units) for electric dipole (E1), and 
A$_{ji}$ for electric quadrupole (E2), magnetic dipole (M1), and magnetic quadrupole (M2) transitions of Ar~X.
  The ratio R(E1) of velocity and length forms of A-values for E1 transitions is listed in the last column.}
\begin{tabular}{@{}p{1in}p{6in}@{}}
$i$ and $j$         & The lower ($i$) and upper ($j$) levels of a transition as defined in Table 7.\\
$\lambda_{ij}$      & Transition wavelength (in ${\rm \AA}$) \\
A$^{E1}_{ji}$       & Radiative transition probability (in s$^{-1}$) for the E1 transitions \\
f$^{E1}_{ij}$       & Absorption oscillator strength (dimensionless) for the E1 transitions \\
S$^{E1}$            & Line strength in atomic unit (a.u.), 1 a.u. = 6.460$\times$10$^{-36}$ cm$^2$ esu$^2$ for the E1 transitions \\
A$^{E2}_{ji}$       & Radiative transition probability (in s$^{-1}$) for the E2 transitions \\
A$^{M1}_{ji}$       & Radiative transition probability (in s$^{-1}$) for the M1 transitions \\
A$^{M2}_{ji}$       & Radiative transition probability (in s$^{-1}$) for the M2 transitions \\
R(E1)                     & Ratio of velocity and length forms of A- (or f- and S-) values for the E1 transitions \\
$a{\pm}b$ &  $\equiv a\times{10^{{\pm}b}}$ \\
\end{tabular}
\label{ExplTable18}

\bigskip
\section*{Table 19.\label{tbl19te}  Transition wavelengths ($\lambda_{ij}$ in $\rm \AA$), radiative rates (A$_{ji}$ in s$^{-1}$),
 oscillator strengths (f$_{ij}$, dimensionless), and line strengths (S, in atomic units) for electric dipole (E1), and 
A$_{ji}$ for electric quadrupole (E2), magnetic dipole (M1), and magnetic quadrupole (M2) transitions of K~XI
 The ratio R(E1) of velocity and length forms of A-values for E1 transitions is listed in the last column.}
\begin{tabular}{@{}p{1in}p{6in}@{}}
$i$ and $j$         & The lower ($i$) and upper ($j$) levels of a transition as defined in Table 8.\\
$\lambda_{ij}$      & Transition wavelength (in ${\rm \AA}$) \\
A$^{E1}_{ji}$       & Radiative transition probability (in s$^{-1}$) for the E1 transitions \\
f$^{E1}_{ij}$       & Absorption oscillator strength (dimensionless) for the E1 transitions \\
S$^{E1}$            & Line strength in atomic unit (a.u.), 1 a.u. = 6.460$\times$10$^{-36}$ cm$^2$ esu$^2$ for the E1 transitions \\
A$^{E2}_{ji}$       & Radiative transition probability (in s$^{-1}$) for the E2 transitions \\
A$^{M1}_{ji}$       & Radiative transition probability (in s$^{-1}$) for the M1 transitions \\
A$^{M2}_{ji}$       & Radiative transition probability (in s$^{-1}$) for the M2 transitions \\
R(E1)                     & Ratio of velocity and length forms of A- (or f- and S-) values for the E1 transitions \\
$a{\pm}b$ &  $\equiv a\times{10^{{\pm}b}}$ \\
\end{tabular}
\label{ExplTable19}

\bigskip
\section*{Table 20.\label{tbl20te}  Transition wavelengths ($\lambda_{ij}$ in $\rm \AA$), radiative rates (A$_{ji}$ in s$^{-1}$),
 oscillator strengths (f$_{ij}$, dimensionless), and line strengths (S, in atomic units) for electric dipole (E1), and 
A$_{ji}$ for electric quadrupole (E2), magnetic dipole (M1), and magnetic quadrupole (M2) transitions of Ca~XII.
The ratio R(E1) of velocity and length forms of A-values for E1 transitions is listed in the last column.}
\begin{tabular}{@{}p{1in}p{6in}@{}}
$i$ and $j$         & The lower ($i$) and upper ($j$) levels of a transition as defined in Table 9.\\
$\lambda_{ij}$      & Transition wavelength (in ${\rm \AA}$) \\
A$^{E1}_{ji}$       & Radiative transition probability (in s$^{-1}$) for the E1 transitions \\
f$^{E1}_{ij}$       & Absorption oscillator strength (dimensionless) for the E1 transitions \\
S$^{E1}$            & Line strength in atomic unit (a.u.), 1 a.u. = 6.460$\times$10$^{-36}$ cm$^2$ esu$^2$ for the E1 transitions \\
A$^{E2}_{ji}$       & Radiative transition probability (in s$^{-1}$) for the E2 transitions \\
A$^{M1}_{ji}$       & Radiative transition probability (in s$^{-1}$) for the M1 transitions \\
A$^{M2}_{ji}$       & Radiative transition probability (in s$^{-1}$) for the M2 transitions \\
R(E1)                     & Ratio of velocity and length forms of A- (or f- and S-) values for the E1 transitions \\
$a{\pm}b$ &  $\equiv a\times{10^{{\pm}b}}$ \\
\end{tabular}
\label{ExplTable20}

\bigskip
\section*{Table 21.\label{tbl21te}  Transition wavelengths ($\lambda_{ij}$ in $\rm \AA$), radiative rates (A$_{ji}$ in s$^{-1}$),
 oscillator strengths (f$_{ij}$, dimensionless), and line strengths (S, in atomic units) for electric dipole (E1), and 
A$_{ji}$ for electric quadrupole (E2), magnetic dipole (M1), and magnetic quadrupole (M2) transitions of Ti~XIV.
 The ratio R(E1) of velocity and length forms of A-values for E1 transitions is listed in the last column.}
\begin{tabular}{@{}p{1in}p{6in}@{}}
$i$ and $j$         & The lower ($i$) and upper ($j$) levels of a transition as defined in Table 10.\\
$\lambda_{ij}$      & Transition wavelength (in ${\rm \AA}$) \\
A$^{E1}_{ji}$       & Radiative transition probability (in s$^{-1}$) for the E1 transitions \\
f$^{E1}_{ij}$       & Absorption oscillator strength (dimensionless) for the E1 transitions \\
S$^{E1}$            & Line strength in atomic unit (a.u.), 1 a.u. = 6.460$\times$10$^{-36}$ cm$^2$ esu$^2$ for the E1 transitions \\
A$^{E2}_{ji}$       & Radiative transition probability (in s$^{-1}$) for the E2 transitions \\
A$^{M1}_{ji}$       & Radiative transition probability (in s$^{-1}$) for the M1 transitions \\
A$^{M2}_{ji}$       & Radiative transition probability (in s$^{-1}$) for the M2 transitions \\
R(E1)                     & Ratio of velocity and length forms of A- (or f- and S-) values for the E1 transitions \\
$a{\pm}b$ &  $\equiv a\times{10^{{\pm}b}}$ \\
\end{tabular}
\label{ExplTable21}

\bigskip
\section*{Table 22.\label{tbl22te}  Transition wavelengths ($\lambda_{ij}$ in $\rm \AA$), radiative rates (A$_{ji}$ in s$^{-1}$),
 oscillator strengths (f$_{ij}$, dimensionless), and line strengths (S, in atomic units) for electric dipole (E1), and 
A$_{ji}$ for electric quadrupole (E2), magnetic dipole (M1), and magnetic quadrupole (M2) transitions of V~XV.
 The ratio R(E1) of velocity and length forms of A-values for E1 transitions is listed in the last column.}
\begin{tabular}{@{}p{1in}p{6in}@{}}
$i$ and $j$         & The lower ($i$) and upper ($j$) levels of a transition as defined in Table 11.\\
$\lambda_{ij}$      & Transition wavelength (in ${\rm \AA}$) \\
A$^{E1}_{ji}$       & Radiative transition probability (in s$^{-1}$) for the E1 transitions \\
f$^{E1}_{ij}$       & Absorption oscillator strength (dimensionless) for the E1 transitions \\
S$^{E1}$            & Line strength in atomic unit (a.u.), 1 a.u. = 6.460$\times$10$^{-36}$ cm$^2$ esu$^2$ for the E1 transitions \\
A$^{E2}_{ji}$       & Radiative transition probability (in s$^{-1}$) for the E2 transitions \\
A$^{M1}_{ji}$       & Radiative transition probability (in s$^{-1}$) for the M1 transitions \\
A$^{M2}_{ji}$       & Radiative transition probability (in s$^{-1}$) for the M2 transitions \\
R(E1)                     & Ratio of velocity and length forms of A- (or f- and S-) values for the E1 transitions \\
$a{\pm}b$ &  $\equiv a\times{10^{{\pm}b}}$ \\
\end{tabular}
\label{ExplTable22}


\section*{References}

\begin{thebibliography}{999}
\bibitem{samp}   D.H. Sampson, H.L. Zhang, C.J. Fontes, At.  Data Nucl. Data Tables  48 (1991) 25.
\bibitem{kr}    K.M.  Aggarwal, F.P. Keenan,  K.D. Lawson,   At. Data Nucl. Data Tables   94 (2008) 323.
\bibitem{xe}   K.M.  Aggarwal, F.P.  Keenan,  K.D. Lawson,   At. Data Nucl. Data Tables   96 (2010) 123.
\bibitem{w66a}   K.M. Aggarwal,  F.P. Keenan, At. Data Nucl. Data Tables  111-112 (2016) 187.
\bibitem{w66b}   K.M.  Aggarwal, Chin. Phys. B 25 (2016) 043201.
\bibitem{w66c}   K.M.  Aggarwal, Atoms 4 (2016) 4030024.
\bibitem{ak1}    K.M.  Aggarwal, F.P. Keenan,    At. Data Nucl. Data Tables   109-110 (2016) 205.
\bibitem{ak2}    K.M.  Aggarwal,    At. Data Nucl. Data Tables   123-124 (2018) 168.
\bibitem{si}       R. Si, S. Li, X.L. Guo, Z.B. Chen, T. Brage, P. J\"{o}nsson, K. Wang, J,. Yan, C.Y. Chen, Y.M. Zou, Astrophys. J. Suppl. 227 (2016) 16.
\bibitem{li}        J.Q. Li, C.Y. Zhang, R. Si, K. Wang, C.Y. Chen, At. Data Nucl. Data Tables   126 (2019) in press.
\bibitem{grasp0} I.P. Grant, B.J.  McKenzie, P.H. Norrington, D.F. Mayers,   N.C. Pyper, Comput. Phys. Commun.   21  (1980) 207.
\bibitem{fac}   M.F.  Gu,  Can. J. Phys.  86 (2008) 675.
\bibitem{gu}   M.F.  Gu,   At. Data Nucl. Data Tables 89 (2005) 267.
\bibitem{jag}  P. J\"{o}nsson, A. Alkauskas, G.  Gaigalas, At.  Data Nucl. Data Tables  99 (2013) 431.
\bibitem{grasp2k} P. J\"{o}nsson, X. He, C.F. Fischer,  I.P. Grant,  Comput. Phys. Commun.  177 (2007) 597.
\bibitem{cff}  C. Froese Fischer, G. Tachiev, At.  Data Nucl. Data Tables  87 (2004) 1.
 \bibitem{team}   J.E. Sansonetti, J.J. Curry,    J. Phys. Chem. Ref. Data 39 (2010)  043103.
\bibitem{sc13} K.M.  Aggarwal, Atoms 6 (2018) 6020025.
\bibitem{jup} C. Jup\'{e}n, E. Tr\"{a}bert, J. Doerfert, J.Granzow,  R. Jaensch, Phys. Scr. 66 (2002) 150.
\bibitem{cow} R.D. Cowan,  ÔÔThe theory of atomic stucture and spectraÕÕ, Univ. of Calif. Press. Berkeley, 1981.
\bibitem{mglike} K.M. Aggarwal,  V. Tayal, G.P. Gupta, F.P. Keenan, At. Data Nucl. Data Tables  93 (2007)  615.
\bibitem{ns} D.K. Nandy, B.K. Sahoo, Astron. Astrophys. 563 (2014) A25.

\end{thebibliography}
\end{document}